# Excitonic Transport and Intervalley Scattering in Exfoliated MoSe$_2$ Monolayer Revealed by Four-Wave-Mixing Transient Grating Spectroscopy

*Henning Kuhn,[†] Julian Wagner,[†] Shuangping Han,[‡, §] Robin Bernhardt,[†] Yan Gao,[∥] Liantuan Xiao,[‡, §] Jingyi Zhu,*[,†] and Paul H. M. van Loosdrecht*[,†]*

[†] Universität zu Köln, Ⅱ. Physikalisches Institut, Zülpicher Straße 77, D-50937 Köln, Germany

[‡] State Key Laboratory of Quantum Optics and Quantum Optics Devices, Institute of Laser Spectroscopy, Shanxi University, Taiyuan, Shanxi 030006, China

[§] Collaborative Innovation Center of Extreme Optics, Shanxi University, Taiyuan, Shanxi 030006, China

[∥] Department of Physics, Shanxi Datong University, Datong 037009, China

* jzhu@ph2.uni-koeln.de,   pvl@ph2.uni-koeln.de

**Abstract:** Exciton intervalley scattering, annihilation and relaxation dynamics, and diffusive transport in a monolayer transition metal dichalcogenides are central to the functionality of devices based on them. This motivated us to investigate these properties in exfoliated high-quality monolayer $MoSe_2$ using heterodyned nonlinear four-wave-mixing transient grating spectroscopy. While free exciton excitations are found to be long-lived (~230 ps), an extremely fast intervalley scattering (≤ 120 fs) is observed leading to a negligible valley polarization, consistent with steady state photoluminescence measurements. The exciton population decay shows an appreciable contribution from exciton-exciton annihilation reactions with an annihilation constant of ~ 0.01 $cm^2s^{-1}$, which in addition leads to an extra contribution to the transient grating response. The underlying excitonic dynamics were numerically modeled by including exciton-exciton annihilation, also in the diffusion equation, which allows extraction of the diffusion constant, D ~1.4 $cm^2s^{-1}$. Our results provide a method that allows for the disentanglement of the intricate dynamics involving many-body annihilation processes and a detailed characterization of the excitonic properties of monolayer $MoSe_2$.

**Introduction:**

Monolayer transition metal dichalcogenides (TMDCs) are a novel type of two-dimensional materials which, following the graphene upsurge[1-3], are now taking central stage in the research of functional two dimensional materials[4-10]. In contrast to graphene, monolayer TMDCs display a variety of intriguing optical and physical properties emerging from the lack of inversion symmetry, reduction of dimensionality, strong excitonic effects[11, 12], and strong spin-orbit interactions. These properties lead to the opening of a direct band gap[4, 13], spin-valley-dependent optical transitions,[8, 14] and the valley Hall effect,[15, 16] and make TMDCs promising candidate materials for photonic and optoelectronic applications[17-22] utilizing the valley polarization in the so-called valleytronics which has the potential to be integrated with spintronics[23, 24] yielding multi-valley material applications.

One of the most prominent properties in monolayer TMDCs are strongly enhanced optical transition at the band gap edges, the fundamental electronic excitations forming excitons, *i.e.* strongly Coulomb-bound electron-hole pairs. Due to the reduced electronic screening and the heavy effective mass of carriers, the exciton binding energy in TMDCs can reach several hundreds of meV,[25, 26] which dominates the optical spectra even at room temperature. Since the optical performance depends crucially on the exciton relaxation dynamics, a detailed understanding of the dynamical behavior of the excitons after photo-excitation is pivotal for the development of applications in the field of optical and optoelectronic devices. Indeed, optical excitation properties and the associated relaxation dynamics in TMDCs have been extensively studied in the recent past using various ultrafast spectroscopic techniques. These studies revealed many interesting and special properties, such as the optically induced giant band gap renormalization,[27, 28] the strong valley-selective optical Stark effect[29-31] and strong sample and excitation dependence of the exciton relaxation dynamics originating from both first and second order relaxation processes[32-35].

Although a variety of experimental and theoretical studies have been carried out to investigate the intrinsic exciton population dynamics in monolayer TMDCs, the transport properties of the optical excitations received less attention[36-38] and provided a range of values for the diffusion constant without a consensus interpretation. Another intensively investigated property of monolayer TMDCs is the valley depolarization and the associated intervalley scattering dynamics. An often used method to study the valley polarization focuses on the steady state through polarized photoluminescence (PL) experiments.[7, 8, 39] The intrinsic femtosecond and picosecond dynamics of the optically excited excitons has, however, received less attention, most likely due to the to lack of efficient and proper methods to study this. For example in monolayer WS$_2$, a very fast valley scattering (few ps[27, 40]) has been reported which is inconsistent with the large degree of valley polarization observed in steady state experiments[41-43]. The reason for this lies in intense optical excitation used in the time resolved experiments which cause a giant band gap renormalization with likely associated changes in the valley potential and the intervalley scattering processes. Very recently, in a chemical vapor deposition (CVD) grown monolayer MoSe$_2$ sample, it was found that the valley polarization is relatively large and depolarization dynamics is excitation dependent.[44] The valley depolarization mechanism was explained in terms of simultaneous exchange of two excitons between valleys. However, this again seems to be in contrast to the nearly zero valley polarization observed for the exfoliated monolayer MoSe$_2$, even at very low temperatures,[45, 46] indicating a more complicated situation in the monolayer MoSe$_2$.

To resolve these issues we applied nonlinear heterodyned four-wave-mixing transient grating (HFWMTG) spectroscopy technique, combined with steady state polarized luminescence spectroscopy to exfoliated monolayer MoSe$_2$. The HFWMTG technique allows determining the exciton-exciton annihilation rate, exciton free decay, and exciton diffusion dynamics simultaneously. Further, a special configuration of the HFWMTG polarization alignment of the laser beams allows investigating the valley scattering dynamics directly. It is found that exciton-exciton annihilation processes not only strongly influence the population dynamics, but also has a profound effect on the HFWMTG dynamics, as it distorts the transient grating shape. To reliably extract the exciton dynamics and the diffusion coefficient we provide a general method to disentangle the dynamics in a system with strong many-body reaction decay processes such as the monolayer TMDCs. Except for extracting the exciton diffusion constant, furthermore we show that even though the excitons have a relatively long lifetime, the valley polarization dynamics is extremely fast. This is in full agreement with the fact that PL experiments on exfoliated MoSe$_2$ show a nearly zero steady state valley polarization.

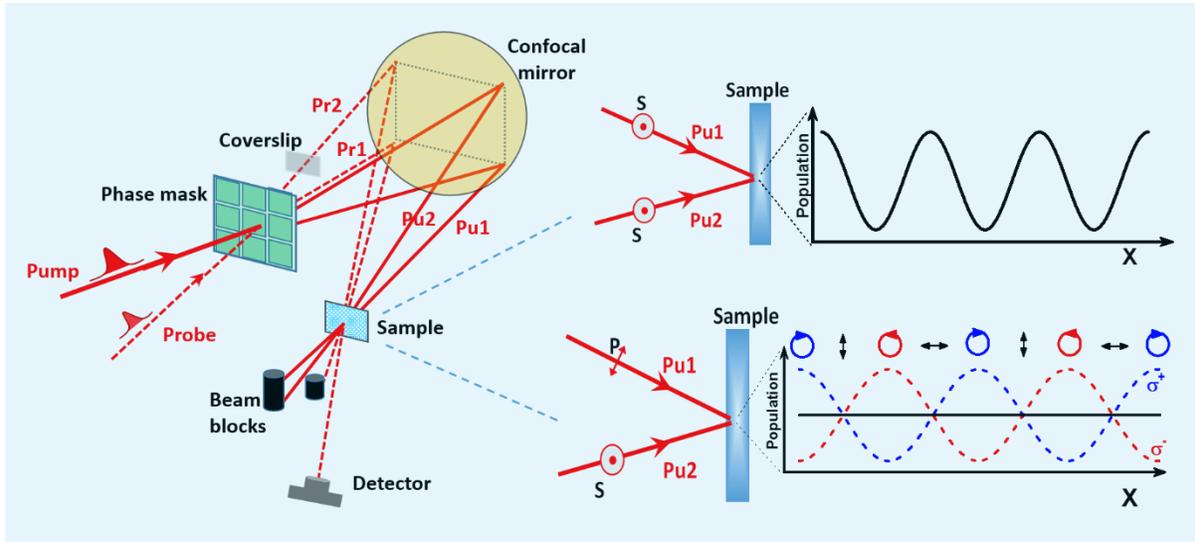

Figure 1. Scheme of the key part of the four-wave-mixing heterodyned transient grating spectroscopy method using a BOXCAR geometry. Laser beams Pu1 and Pu2 act as excitation, laser beams Pr1 and Pr2 act as probe. The detector is placed in the propagation direction of beam Pr2. In this detection direction, beam Pr2 acts as both the heterodyne pulse for monitoring the grating signal resulting from interfering excitation by Pu1 and Pu2 and diffraction of Pr1, and the probe of the population signals from each pump of Pu1 and Pu2, respectively. The relative optical phase between Pr1 and Pr2 was controlled by a rotatable coverslip. When the polarization of beam Pu1 and Pu2 are set parallel, a population grating is created, when polarizations of beam Pu1 and Pu2 are set vertical to each other, a spin or valley grating is created.

**Experimental methods:**

Monolayer MoSe$_2$ flakes were obtained by exfoliation with adhesive tape from a bulk single crystal and deposited on a c-cut sapphire substrate. First, a large area monolayer MoSe$_2$ was identified along with the multilayers by imaging with an optical microscope utilizing the large optical contrast. Further characterization of the MoSe$_2$ flake was performed using steady state PL and Raman scattering spectroscopy. Raman and PL spectra were measured using a micro-Raman setup equipped with a triple stage spectrometer (Spectroscopy & Imaging GmbH) and a liquid nitrogen cooled (-120 °C) CCD detector (PyLoN 100; Princeton Instruments). PL spectra were obtained with a continuum laser excitation at either 532 nm or 780 nm (for valley polarization), and Raman spectra were measured using a continuum laser with center wavelength set to 532 nm. The laser pulses were focused on the sample using a microscope objective (20×, NA=0.4). Raman and PL signals were collected in a backscattering geometry.

Ultrafast exciton relaxation and transport dynamics in the monolayer MoSe$_2$ were measured simultaneously by HFWMTG. In the HFWMTG technique (Figure 1), the standard BOXCAR configuration[47, 48] of the optical path alignment was employed. Two parallel propagating Gaussian laser beams (pump and probe) derived from the same laser source (Mira 900-F, Coherent) were focused on a phase mask (custom made transmission grating) which replicates the two incoming beams into four (Pu1, Pu1, Pr1, and Pr2, from each beam's positive and negative first order diffraction). A spherical confocal mirror focuses the incoming four beams onto the sample, in which the intense replica (Pu1 and Pu2) act as excitation, to create an interference grating pattern along the sample. The spatially modulated excitation pattern is probed by the two weak probe beams Pr1 and Pr2. Beam Pr1 gets diffracted by the excitation induced grating in the sample, forming the transient grating signal in the sense of third order nonlinear optics. Due to the special geometry of the BOXCAR configuration, the transient grating signal follows exactly the beam direction of the specular reflected beam Pr2, which acts as a phase sensitive amplifying source to the signal and the phase can be adjusted by the angle of a microscope cover slip in its beam path, i.e. heterodyne detection. The heterodyned signal is then sent to a lock-in amplifier which detects the probe signal at the 100 kHz frequency of the chopper (photoelastic modulator, PEM-90, Hinds Instruments) in the pump path. The laser has a repetition rate of 75 MHz, a pulse duration of ~200 fs. To resonantly excite monolayer $MoSe_2$ and minimize laser induced heating, the laser wavelength was tuned to 750 nm (~1.6 eV), corresponding to the lowest excitonic transition of monolayer $MoSe_2$. Note that in the BOXCAR configuration, the detected signal also includes other two self-phase matching third order nonlinear components as the normal pump probe measurements, i.e. signals from Pu1 pump Pr2 probe and Pu2 pump Pr2 probe. These signals contain the same information and represent the transient transmission signals of the population decay dynamics only. To obtain both the transport and population dynamics simultaneously and in order to separate them, two-dimensional data traces were recorded by controlling both the time delay between the pump and probe (delay line control) and the optical phase difference between the both probe beams. For the population grating, both of the excitation beams (Pu1 and Pu2) are parallel linear polarized. While for the spin grating or valley grating,[44, 49, 50] cross-polarized excitation laser beams were applied to generate a spatially modulated distribution of circularly polarized states with uniform intensity. To probe the valley grating, the two probe beams (Pr1 and Pr2) are also linearly cross-polarized to each other.[44]

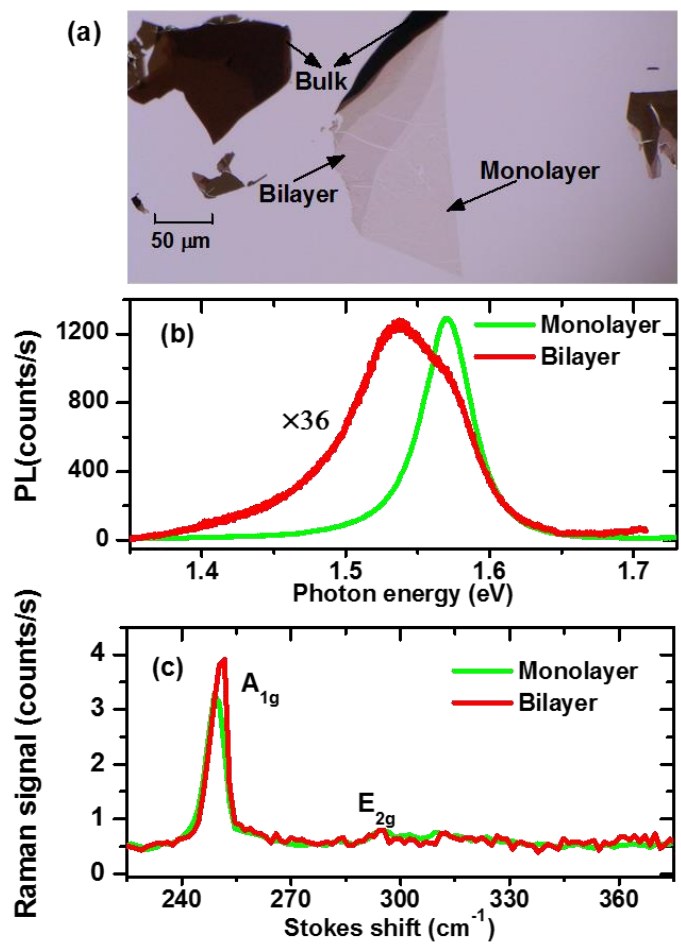

Figure 2. Characterization of the mechanically exfoliated large size MoSe$_2$ flakes on a c-cut sapphire substrate. (a) Optical microscope image. (b) Photoluminescence and (c) Raman spectra of monolayer and bilayer MoSe$_2$. The spectra were recorded with a continuum laser at 532 nm, using an excitation effective power of ~10 μW at the sample position and with a beam spot diameter of ~ 3 μm.

**Results:**

Figure 2 (a) shows the optical microscopy image of large size exfoliated flakes of the MoSe$_2$ on a c-cut sapphire substrate. The monolayer region obtained here has a size of ~100 μm × 50 μm, and is cling by a relatively smaller size of bilayer and other multilayers. The layer properties of the mechanically exfoliated flakes were further examined by the steady state PL and Raman spectra, as presented in Figure 2 (b) and (c), respectively. The PL spectrum of the monolayer $MoSe_2$ has a peak centered at ~ 1.57 eV and a peak width (full width at the half maximum, FWHM) of ~ 40 meV, corresponding to the lowest excitonic transition. The bilayer PL spectrum shows a broadened and red shifted peak centered at ~ 1.54 eV. These PL results are consistent with previous reports for mechanically exfoliated $MoSe_2$.[51, 52] Note that the obtained PL signal for the bilayer is ~ 36 times weaker than the PL spectrum of the monolayer under the same measurement conditions, indicating an indirect-to-direct band gap transition from bilayer to monolayer. In the measured Raman spectra, both monolayer and bilayer exhibit very similar features, as both have two major peaks, one located at around 251 $cm^{-1}$, ascribed to the $A_{1g}$ optical phonon mode, and the other located at around 294 $cm^{-1}$, ascribed to the $E_{2g}$ phonon mode.[51, 53-55] Comparing to the $A_{1g}$ mode peaks, the $E_{2g}$ peaks are much weaker in both the monolayer and bilayer. The $A_{1g}$ mode peak in the bilayer is slightly stronger than that in monolayer. Although the difference in Raman spectra cannot easily be used alone as a discrimination of number of layers, these observations of Raman scattering coincide with previously reported results of Raman spectra of layered $MoSe_2$.[51, 53, 55]

Typical heterodyned population transient grating signals for the monolayer $MoSe_2$ are presented in figure 3, at an excitation intensity of ~13.8 $\mu Jcm^{-2}$ (corresponding to an exciton density ~2.6 × $10^{12} cm^{-2}$). For detection of the population related dynamics, the polarizations of the two pump beams are set parallel to each other. Figure 3 (a)-(c) present a few representative examples of the measured two-dimensional data for varying both the pump-probe delay and the optical phase difference of the heterodyne pulse. As it can been seen clearly, in the first tens of ps time range, a slight phase shifting of the grating signal occurs, indicating the necessity of the two dimensional recording in order to guarantee the precise measurement of the grating dynamics. From these two dimensional transients data, population dynamics (constant offset level of the sinusoidal signal) and grating dynamics (amplitude of the sinusoidal signal) can be conveniently extracted[56] by a routine fitting procedure. The extracted population ($S_{pop}$) and grating ($S_{GR}$) transients are presented in figure 3 (d) and (e), respectively. Here, $S_{pop}$ represents the pure exciton population dynamics while $S_{GR}$ contains both information of the population and diffusion dynamics. Apparently, the grating signal $S_{GR}$ in figure 3(e) decays much faster than that of the population in figure 3(d), indicating a significant diffusion contribution to the dynamics in the exfoliated monolayer $MoSe_2$. This is a very different observation compared to the results obtained for the CVD monolayer $MoSe_2$[44] where the excitation dynamics show no diffusive contribution at all.

Before fully analyzing the exciton population and diffusion dynamics, the excitation density effects on both of the population and grating dynamics were examined by varying the pump laser intensities over an order of magnitude. Figure 4 (a) shows the extracted population dynamics as treated in figure 3 (d) but at different laser pump intensities. As the pump intensity increases, the population decay is clearly speeding up. Such an observation is again in contrast to the situation of the CVD grown monolayer $MoSe_2$, in which the population decay has no excitation density dependence[44]. This indicates that in the exfoliated monolayer $MoSe_2$, besides the free exciton recombination dynamics, a many-body reaction is involved, i.e. the exciton-exciton annihilation process plays an important role in the exciton population dynamics.

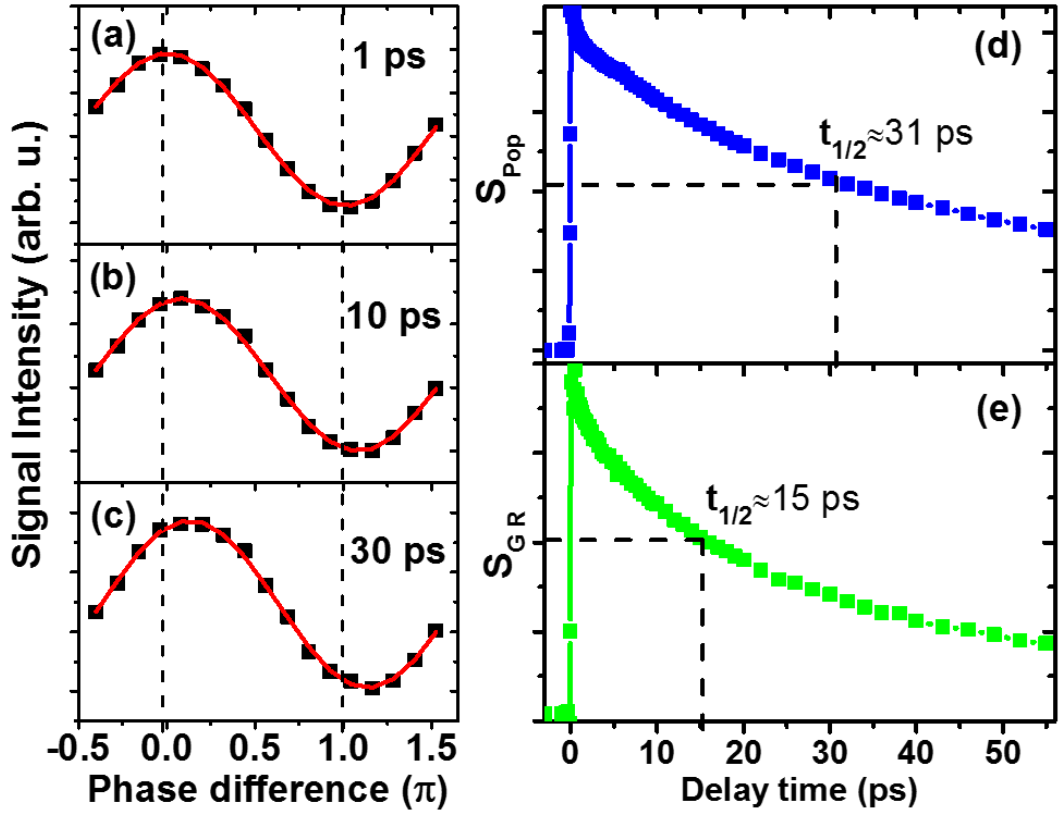

Figure 3. Population and transient grating signals of monolayer MoSe$_2$ extracted from the two-dimensional data traces of the heterodyne detection at an excitation intensity of ~13.8 μJcm$^{-2}$ (~2.6 × 10$^{12}$cm$^{-2}$) for a grating period of $\Lambda_{TG}$=2 μm. (a)-(c) Signals as a function of the heterodyne phase difference between two probe pulses scanned at different delay times. Black dots show the experimental data, red curves correspond to a sinusoidal fit. (d) Extracted population dynamics. (e) Extracted transient grating dynamics. Dashed lines in (a)-(c) indicate the phase shift and in (d)-(e) indicate the half lifetimes of the decay dynamics.

To analyze these observed excitation density dependent dynamics more conveniently, we simply expressed (see supplementary information) the time dependent population density N(t) as:

$$N(0)/N(t) - 1 = [k_1 + N(0) k_2]t \qquad (1)$$

Here N(0) is the initial exciton density created by the pump pulses at zero delay time, $k_1$ and $k_2$ are the free exciton decay and exciton-exciton annihilation rates, respectively. Equation (1) presents a very compact and intuitive description of the population dynamics: the ratio $(N(0)-N(t))/N(t)$ is simply linearly proportional to the delay time t, while the slope is dependent on the initial excitation density if second order recombination dynamics plays a role. Indeed, as shown in figure 4 (b), this is exactly what is observed for the density dependent dynamics. Global fitting (green lines) of the data (symbols) using formula (1) yields satisfying agreement, yielding a first order rate constant $k_1 = (230\ ps)^{-1}$, and a second order constant $k_2 = 0.01\ cm^2 s^{-1}$.

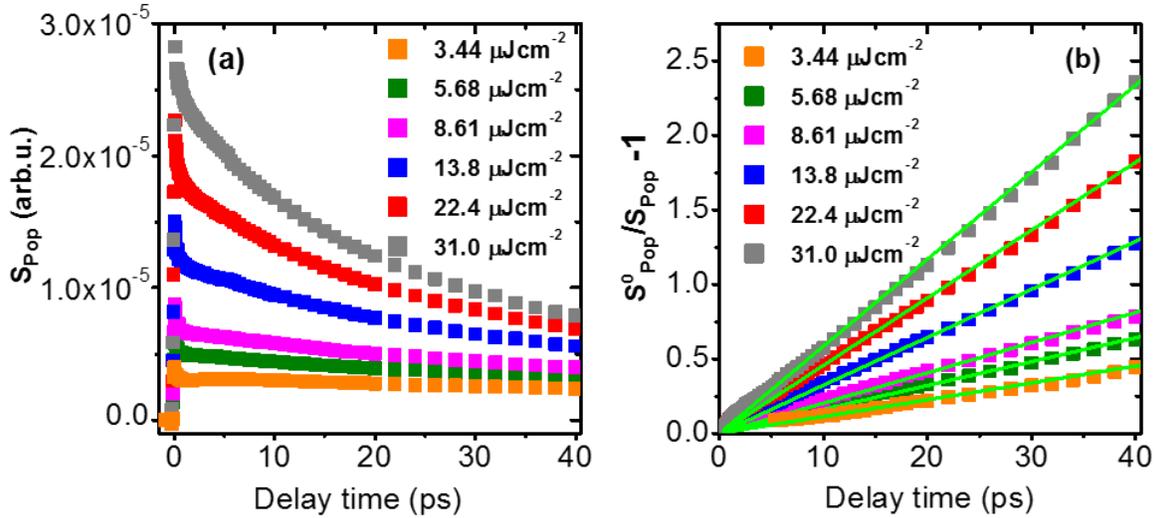

Figure 4. Excitation intensity dependent exciton population dynamics in exfoliated monolayer MoSe$_2$. (a) Raw data. (b) Normalized data $(S_{pop}(0)-S_{pop}(t))/S_{pop}(t)$ showing a linear dependence on the delay time. Dots are experimental data and lines are obtained from a global fit. Fitted free exciton decay rate $k_1 = (230\ ps)^{-1}$ and exciton-exciton annihilation constant $k_2 = 0.01\ cm^2 s^{-1}$ as indicated in the inset of (b).

Conventionally, grating dynamics are discussed using first order dynamics only. Figure 5 (a) (squared dots), however, shows that also the grating-signal decay is accelerated with increasing excitation density. This acceleration is caused by two aspects: the exciton-exciton annihilation dependent population dynamics, and the inhomogeneous distortion of the grating pattern due to spatial-intensity-dependent annihilation which influences the grating diffraction efficiency. For the latter, specific model simulations and the derivation of new formulas are in order, which will be addressed in detail in the discussion part.

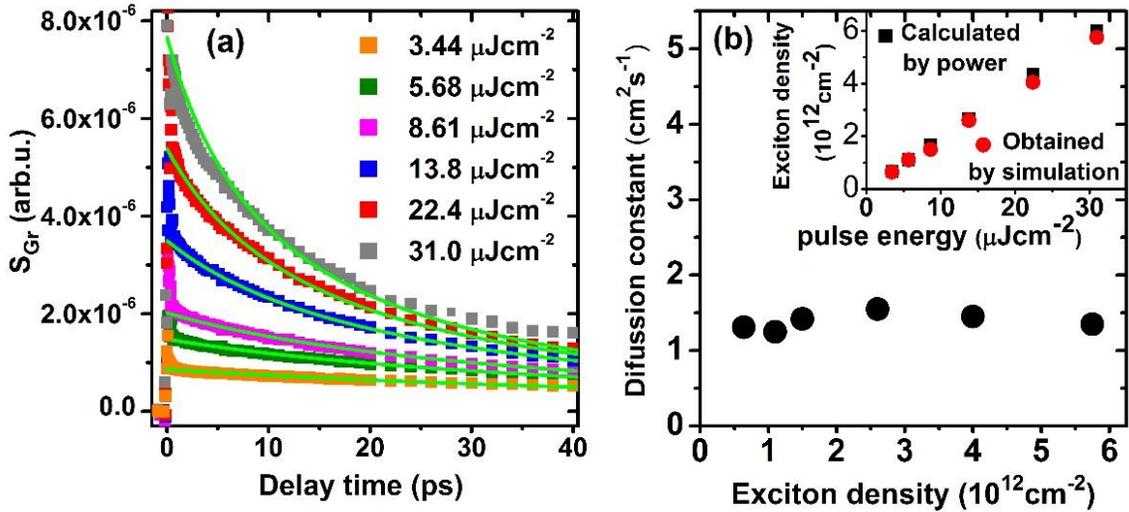

Figure 5. Transient grating signals of monolayer MoSe$_2$ and simulated results with numerical solution of the diffusion equation (3) at different excitation densities. (a) Experimental data (colored dots) and simulated dynamics (green curves). (b) Extracted exciton diffusion constants at different exciton densities revealing a diffusion constant D ~1.4 cm$^2$s$^{-1}$. Inset in (b) indicates the exciton densities obtained from the laser parameters calibration and this simulation.

Lastly we interrogated the valley depolarization dynamics in the exfoliated monolayer MoSe$_2$. To perform these measurements, the electric field of the two pump (and probe) beams are set to be cross-polarized. As presented in figure 6, the extracted valley grating signals show extremely fast valley depolarization dynamics at both the lower excitation intensity (~ $1.5 \times 10^{11}$ cm$^{-2}$) and the higher excitation intensity (~$1.6 \times 10^{12}$ cm$^{-2}$) regime. Fitting the data traces with a Gaussian function with a FWHM of ~170 fs shows consistency with the decay dynamics for both exciton densities. The fitted FWHM matches the autocorrelation pulse width of our laser pulse (FWHM ~ 120 fs), indicating that the valley polarization decay dynamics is extremely fast, below the time resolution of our setup.

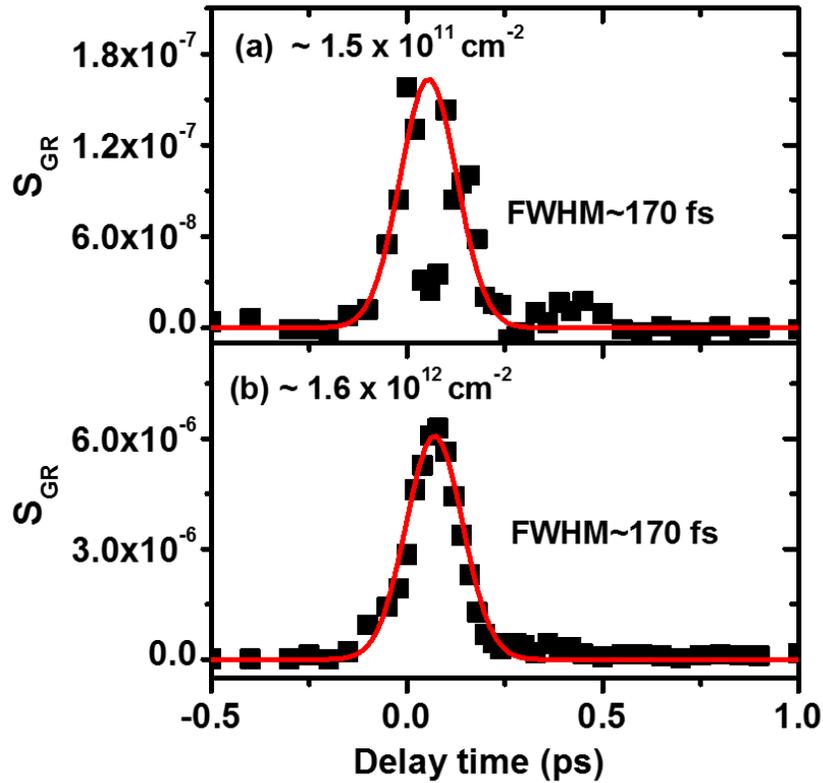

Figure 6. Valley depolarization dynamics in exfoliated monolayer MoSe$_2$ at different excitation densities. (a) At the excitation density of ~ $1.5 \times 10^{11}$ cm$^{-2}$. (b) At the excitation density of ~ $1.6 \times 10^{12}$ cm$^{-2}$. Dots display the extracted grating decay signal from experiments and red lines are Gaussian fits with a full width at half maximum of ~ 170 fs.

**Discussion:**

In the general case, where the grating dynamics only involves the free particle decay, the relation between grating decay rate and the population decay rate can be simply described by[57-59]:

$$\tau^{-1}_G = \tau^{-1}_P + q^2 D \tag{2}$$

where $\tau_G$ and $\tau_P$ are the observed grating and population decay lifetime constants, q is the grating wave vector which is proportional to the inverse of the grating period (q=2π/$\Lambda_{TG}$, $\Lambda_{TG}$ is the grating period created in the sample), and D is the diffusion constant. This relation can be readily derived under the assumption that the grating decay is uniform without any spatial distortion during the time evolution. This assumption holds if there is only first order decay of free particles. However, if there are second order decay channels, i.e. here the exciton-exciton annihilation dynamics, such processes will induce spatially dependent recombination which will introduce distortion of the grating pattern with time. To describe the dynamics quantitatively and to understand the grating evolution more clearly, we include the exciton-exciton annihilation term in the diffusion equation:

$$\frac{\partial N(x,t)}{\partial t} = D \frac{\partial^2 N(x,t)}{\partial x^2} - k_1 N(x,t) - k_2 N^2(x,t) \tag{3}$$

and numerically solve it for given parameters D, $k_1$ and $k_2$ and initial sinusoidal grating pattern. For an illustration example, a real-time population grating shape evolution with delay time was simulated with presumed parameters and presented in figure 7. As it can be seen, in the first few ps after laser induced excitation, the grating keeps its initial sinusoidal shapes, while for delays larger than 10 ps the spatial distortion becomes clearly visible, in which the peaks are suppressed and broadened, while the valleys get relatively narrowed.

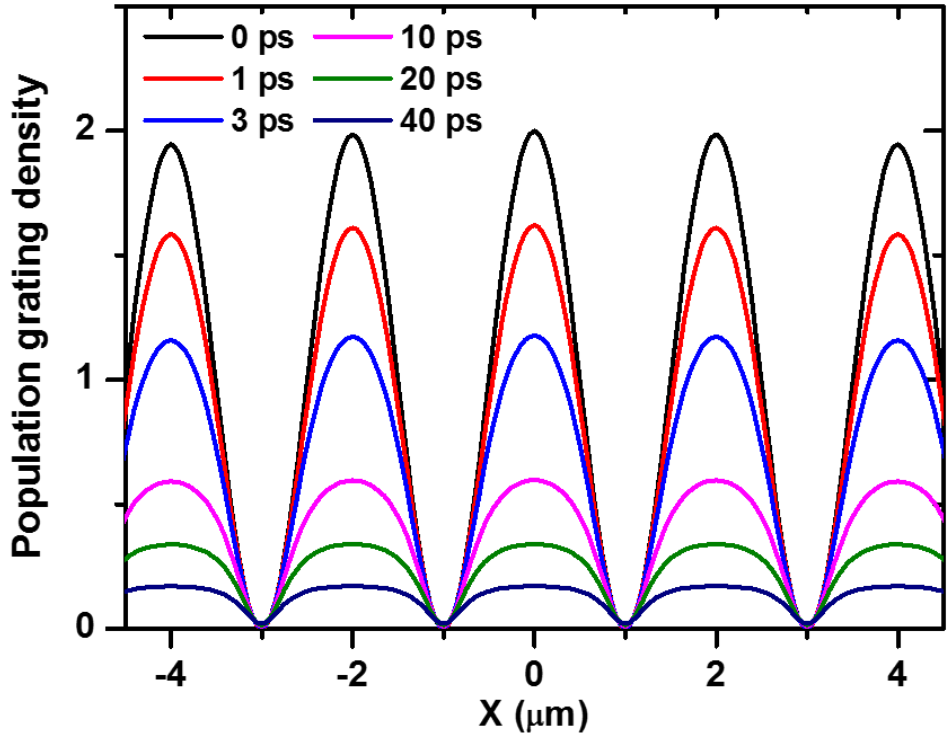

Figure 7. Demonstration plots of a numerically solved transient grating pattern at different delay times including both the free exciton decay and the second order exciton-exciton annihilation processes. The initial sinusoidal grating is distorted with delay time due to the second order reaction and exciton diffusion. The diffracted signal from this kind of grating can be calculated based on physical models (see supplementary). Parameters to generate these gratings are: free exciton decay rate $k_1 = (100\text{ ps})^{-1}$, exciton-exciton annihilation rate $k_2 = 1\text{ cm}^2\text{s}^{-1}$, diffusion constant $D = 1\text{ cm}^2\text{s}^{-1}$ and grating period $\Lambda_{TG} = 2\text{ μm}$. Temporal evolution of the grating pattern was obtained by solving the diffusion equation (3) numerically.

The grating distortion induces changes in the diffraction efficiency, and hence in the grating signals. Based on a simple absorptive material model, the grating signal arising from the distorted grating can be written as (see supplementary):

$$S_{TG}(t) \propto \int I_0(x)N(x,t)\cos(qx)dx \qquad (4)$$

in which $I_0(x)$ is the spatial profile of the probe laser beam, and $N(x,t)$ is the distorted population grating pattern in space and time, which can be obtained by solving equation (3) numerically. By doing this, the experimentally extracted grating decay data at different excitation densities can be well simulated, as shown by the green lines in figure 5 (a), leading to a excitation density independent diffusion constant $D \sim 1.4$ cm$^2$s$^{-1}$, as presented in figure 5(b). We also compared the initial populations at different excitation intensities resulting from the simulations to those obtained by estimating the laser induced exciton density from the pulse energy of the excitation laser pulses and the material absorption parameters. The good agreement found (see inset figure 5b) gives further validation of the model used.

Although, exciton-exciton annihilation processes are commonly observed phenomena in many crystalline materials[60-64], the first observation of this phenomenon in monolayer TMDCs is in an exfoliated monolayer MoS$_2$, where a moderate annihilation constant of 0.04 cm$^2$s$^{-1}$ was found.[32] Later on in different samples of monolayer WS$_2$, conflicting and confusing results have been reported, including statements that the annihilation process is absent[65], is highly efficient[33] with a huge annihilation constant of 0.41 cm$^2$s$^{-1}$, or is less efficient with annihilation constant around 0.1 cm$^2$s$^{-1}$.[34, 35] These inconsistent results may be understood by the different defect densities in specific samples. In samples with a larger defect density, excitons would be less likely to meet each other, resulting in a smaller value of the annihilation rate. Here the observed annihilation rate value ($\sim 0.01$ cm$^2$s$^{-1}$) in the exfoliated MoSe$_2$ of our sample is smaller than that of an exfoliated monolayer MoS$_2$,[32] which is reasonable considering the much easier oxidation of selenide comparing to sulfide, which might cause more oxidation defects in MoSe$_2$.

The obtained diffusion constant $D = 1.4$ cm$^2$s$^{-1}$ is much smaller than the previously reported result in the exfoliated monolayer MoSe$_2$.[37] Similarly to the exciton-exciton annihilation rate, a possible reason is that the exfoliation process unintentionally introduced defects are different. Perhaps there are more vacancies in our sample, which lead to a higher trapping probability and hence limit the diffusion. However, such a large difference more likely has its origin in the applied measurement method and the significant contributions of second order exciton annihilation to the dynamics. Using the obtained diffusion constant of 1.4 cm$^2$s$^{-1}$, we can estimate the exciton effective mass $M_{ex}$ using the Einstein relation: $D \approx k_B T \tau / M_{ex}$, where $k_B$ is the Boltzmann constant, T is temperature (~300 K) and $\tau$ is the exciton de-coherence time (~34 fs as follows from the PL experiments, see supplementary). This yields an effective exciton mass of 1.1$m_0$ ($m_0$ is the electron rest mass), which is close to value expected from the valence and conduction band dispersions of the K valley of monolayer MoSe$_2$.[66, 67] This agreement further supports our detailed analysis of the HFWMTG experiments and the validity of the obtained rate and diffusion constants. Interestingly, a recent transient grating spectroscopy experiment[38] showed that the exciton-exciton annihilation effect in a high-quality mechanically exfoliated monolayer WSe$_2$ also has huge influence on both the population and grating dynamics. The reported diffusion constant value (D = 0.7 cm$^2$s$^{-1}$), however, is again found to be much smaller than estimated previously from other methods[36, 68] in which exciton-exciton annihilation was not considered.

The observed extremely fast valley depolarization (≤120 fs) indicates extremely small steady state valley polarization at room temperature. Concerning simple rate equations to model the imbalance of the valley excitons the measured steady state valley polarization can be related to the valley scattering and population lifetimes[6, 69]:

$$P = 1/(1+2\ \tau_p/\tau_v) \qquad (5)$$

in which $\tau_p$ and $\tau_v$ are exciton population and valley scattering lifetime. Using the population lifetime $\tau_p$ ~230 ps and valley scattering lifetime of ~120 fs as derived from our experiments, a nearly zero steady state valley polarization degree is obtained, consistent with most of the steady state PL measurements at room temperature, as shown in the supplementary materials and also to previously reported results even at extremely low temperatures.[45, 46] In the past, the exciton intervalley scattering mechanism in TMDCs has been either explained in terms of interaction with longitudinal acoustic phonons[69-71] or as a consequence of the so called electron-hole exchange interaction.[72-75] These two mechanisms can explain observations for most monolayer TMDCs and predict valley scattering lifetimes in the order of a few ps, however exceptions exist for monolayer $MoSe_2$ and $MoTe_2$,[45, 76, 77] which both show negligible steady state circular polarization degree. This conundrum has been solved by a recent model calculation[78] in which a fluctuating Rashba potential is considered. This fluctuating Rashba potential field strongly enhances the mixing of dark and bright exiton states in monolayer $MoSe_2$ due to the very small energy difference[79] between the two states. From the model the calculated intervalley scattering time is less than 1 ps, well in line with our experimental observation.

**Conclusions:**

In conclusion we have systematically investigated the exciton intervalley scattering, and population and transport dynamics in an exfoliated monolayer $MoSe_2$ at different excitation densities using four-wave-mixing heterodyned transient grating spectroscopy. The exciton valley depolarization is found to be extremely fast (≤120 fs) while the exciton population decay is much slower (~230 ps), in line with the absence of valley polarization in steady state PL measurements. The extremely fast intervalley scattering can be understood in terms of a fluctuating Rashba potential induced mixing of dark and bright exciton states. The transient grating dynamics have been found to be strongly affected by second order exciton-exciton annihilation. Using an approach which takes this properly into account we found an excitation density independent diffusion constant of ~1.4 $cm^2s^-$ [1]. Apart from providing detailed knowledge of the exciton dynamics in exfoliated monolayer MoSe$_2$, the present results also provide a method to better describe the transient grating response in systems where many-body annihilation plays an important role.

**Corresponding Authors**

* jzhu@ph2.uni-koeln.de

* pvl@ph2.uni-koeln.de

**ACKNOWLEDGMENT**

This work is partially supported by the DFG (German Research Foundation) via the project No. 277146847 - CRC1238 Control and Dynamics of Quantum Materials, by the National Natural Science Foundation of China (NSFC) [Grant nos. 61605104], and by the Scientific and Technological Innovation Programs of Higher Education Institutions in Shanxi [Grant no. 201802008].

**Supplementary for** 'Excitonic Transport and Intervalley Scattering in Exfoliated MoSe$_2$ Monolayer Revealed by Four-Wave-Mixing Transient Grating Spectroscopy'

**1) Population dynamics including the free exciton decay and exciton-exciton annihilation and the derivation of the linear time dependence:**

The exciton relaxation dynamics can be described by a rate equation

$$\mathrm{d}N(t)/\mathrm{d}t = -k_1 N(t) - k_2 N^2(t),  \quad (s1)$$

where $N(t)$ is the time dependent exciton density, and $k_1$ ($k_2$) is the first (second) order rate constant. Solving equation (s1) analytically gives

$$1/N(t) = [1/N(0) + k_2/k_1]\exp(k_1 t) - k_2/k_1, \qquad (s2)$$

in which $N(0)$ is the initially excited exciton density. Since the effective lifetime of the first order reaction is usually much longer compared to that from the second one, i.e. $k_1 \ll k_2 N(t)$, in this condition and in the short time range, we can safely expand $\exp(k_1 t)$ to $(1 + k_1 t)$ and (s2) becomes:

$$N(0)/N(t) - 1 = [k_1 + N(0)k_2]t \qquad (s3)$$

The right side of (s3) is linearly dependent on the delay time $t$ and the slope constant is excitation density $N(0)$ dependent if the second order annihilation decay play a role.

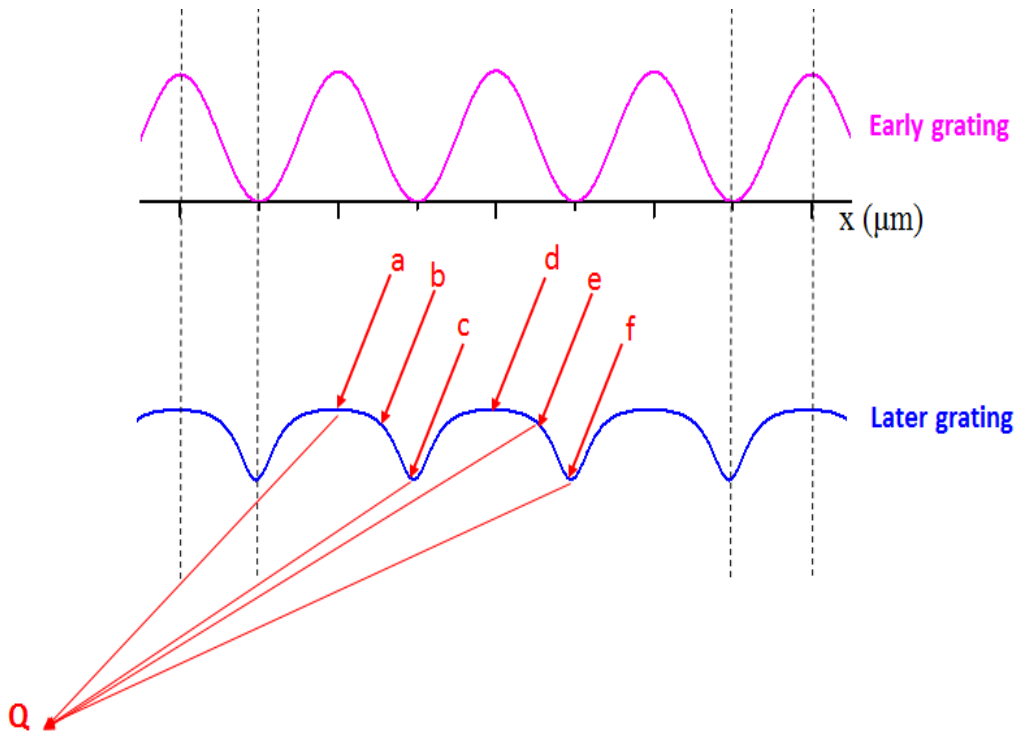

**Figure S1. Diffraction of the probe beam line (a-e) by a transient population grating. Upper and lower panels indicate the initial grating at delay time zero with sinusoidal and later time distorted shapes, respectively. The diffraction signal detected at point Q comes from integrated contributions of all the probe beam lines impinged at different position of the transient grating.**

**2) Modeling of the diffracted signal from a distorted grating**

Here we consider the grating signal, which is from the diffraction of the probe beam, by taking into account optical interaction between the incident probe beam and the sample, i.e. the absorption and reflection effects due to the sample and the substrate. The temporal evolution of the laser induced transient grating is shown in figure S1. At the delay time zero the grating has a sinusoidal shape. With delaying time the grating decays due to the relaxation processes of pump induced excitons and their diffusion along the excitation landscape gradient towards lower excitation areas. This equilibration affects the grating shape in such a way that the grating amplitude shrinks and the sinusoidal shape gets distorted due to the second order relaxation of the exciton-exciton annihilation. However, the phase relation of local grating extrema stays fixed, i.e. the phase factor is the same as for the initial sinusoidal one. Thus, at the detection position Q, it is clear that detraction from light line a and d are in phase, c and f are in phase, while a and c are out of phase. For a random light line e which hits on the grating at position x, the diffraction is proportional to a phase factor $\cos(qx)$, with $q = 2\pi/\Lambda_{TG}$ and $\Lambda_{TG}$ is the grating period in the sample. Therefore, the total diffraction signal at point Q can be expressed as:

$$E_s \propto \int E_s^x \cos(qx) \mathrm{d}x \tag{s4}$$

Here, $E_s^x$ corresponds to a uniform transmitted field amplitude of the probe determined by the material.

In order to estimate $E_s^x$, we consider the approximation of an extreme thin sample, as indicated by Fig. S2. Here, the phase change of the light can thus be ignored. Such a treatment is valid as the thickness of the monolayer sample (few nm) is much smaller compared to the wavelength of the laser beams (750 nm). We also consider the fact that usually a pump induced change of reflection is 1-3 orders smaller than that of the absorption, so we treat the reflection as constant with and without pump. The diffracted light field (proportional to the transmitted filed) is then given by:

$$E_s^x \propto \sqrt{I_0(1-R_1)(1-R_2)\exp(-\alpha L)} \propto \sqrt{I_0}\,(1-0.5\alpha L) \tag{s5}$$

where $I_0$ is the probe beam intensity, $\alpha$ is the absorption coefficient of monolayer $MoSe_2$, and $L$ is the thickness of the monolayer sample. Note that here, we define population dependent $\alpha$ to be

$$\alpha = \sigma\, N_V = \sigma\, N_S\, L^{-1}, \tag{s6}$$

in which, $\sigma$ is the effective absorption cross section, $N_V$ and $N_S$ are unit cell density per volume and per area seen by the probe pulse, i.e. the yet non-excited fraction that gives absorption after the pump. So, we can write

$$N_S = N_0 - N_{exc} \tag{s7}$$

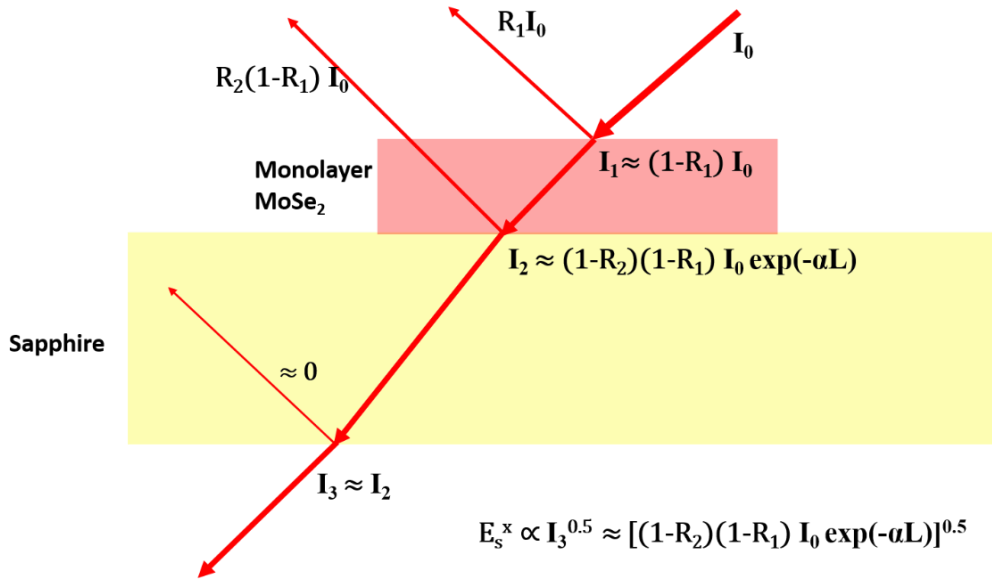

**Figure S2. Transmitted field intensity of the initial probe light intensity $I_0$ after interaction with the monolayer MoSe$_2$ and the sapphire substrate. In the transient grating measurements, the amplitude of the diffracted light field is proportional to the amplitude of the transmitted light field.**

In which $N_0$ is the initial unit cell density without pump, and $N_{exc}$ is the pump excited one, i.e. the transient population part.

Insert (s6) and (s7) into (s5), we have

$$E_s^x \propto \sqrt{I_0}\left(1 - \frac{1}{2}N_0\sigma + \frac{1}{2}N_{exc}\sigma\right) \tag{s8}$$

In addition, we need to take into account the field amplitude of the heterodyne detection beam $E_P$, which is also proportional to $\sqrt{I_0}$, so that the transient grating signal $S_{TG}$ has now to be expressed as:

$$S_{TG} \propto E_s E_p \propto I_0 \int \left(1 - \frac{1}{2}N_0\sigma + \frac{1}{2}N_{exc}\sigma\right)\cos(qx)\mathrm{d}x$$

$$= \tfrac{1}{2}\sigma I_0 \int N_{\text{exc}}\cos(qx)\mathrm{d}x \qquad (s9)$$

The value for $N_{\text{exc}}$ is the numerically determined exciton distribution according to the diffusion equation including the exciton-exciton annihilation, as described in the main text. In this way, the complete dynamics can be simulated to the experimental data, and diffusion constants can be obtained.

Moreover, we can further show that the above derived expression of the transient grating signal is general and can be viewed as an extension to the case with only the free exciton decay included. To do so, we start with the bare one-dimensional diffusion equation but neglect the quadratic term in the excitation density $N_{\text{exc}}$.

$$\frac{\partial N_{\text{exc}}(x,t)}{\partial t} = D\frac{\partial^2 N_{\text{exc}}(x,t)}{\partial x^2} - k_1 N_{\text{exc}}(x,t) \;\cancel{-\; k_2 \Delta N_{\text{exc}}^{\,2}(x,t)} \qquad (s10)$$

and assume the initial excited spatial grating distribution to be:

$$N_{\text{exc}}(x, t=0) = \frac{N_0^{\text{exc}}}{2}[1+\cos(qx)] \qquad (s11)$$

Solving the diffusion equation and respecting the boundary condition for the initial excitation density one finds for the spatio-temporal evolution of the excitation density:

$$N_{\text{exc}}(x,t) = \frac{N_0^{\text{exc}}}{2}\{\exp(-k_1 t) + \cos(qx)\cdot\exp[-(q^2 D + k_1)t]\} \qquad (s12)$$

Here, $q = \frac{2\pi}{\Lambda_{\text{TG}}}$ is the modulus of the grating wave vector and $D$ is the diffusion constant. The first term in the right side of (s12) gives the overall recombinative and spatial independent relaxation like in standard pump-probe experiments and the second term monitors the time evolution of the spatial modulated excitation density, i.e. the transient grating dynamics. Plugging (s12) into the expression (s9) and perform the integration in the symmetrical space region of the transient grating from $-\Delta x$ to $+\Delta x$, where $\Delta x = n \cdot \Lambda_{TG}$ covers an integer number of the grating period $\Lambda_{TG}$, we have:

$$S_{TG}(t) \propto \frac{1}{2}\sigma I_0 \frac{N_0^{exc}}{2} \exp[-(q^2 D + k_1)t] \cdot 2\Delta x$$

$$\propto \exp[-(q^2 D + k_1)t]$$

$$\stackrel{def}{=} \exp\left[-\frac{t}{\tau_{TG}}\right] \tag{s13}$$

In the last step we defined the grating decay rate $\frac{1}{\tau_{TG}} = q^2 D + k_1$. The overall relaxation encoded in the grating signal has two components corresponding to the different decay channels, namely the population decay rate $k_1$ and the diffusive component $q^2 D$. This result is consistent with the standard result from the traditional transient grating principles. [1-4]

**3) Steady state valley polarization of the monolayer MoSe$_2$**

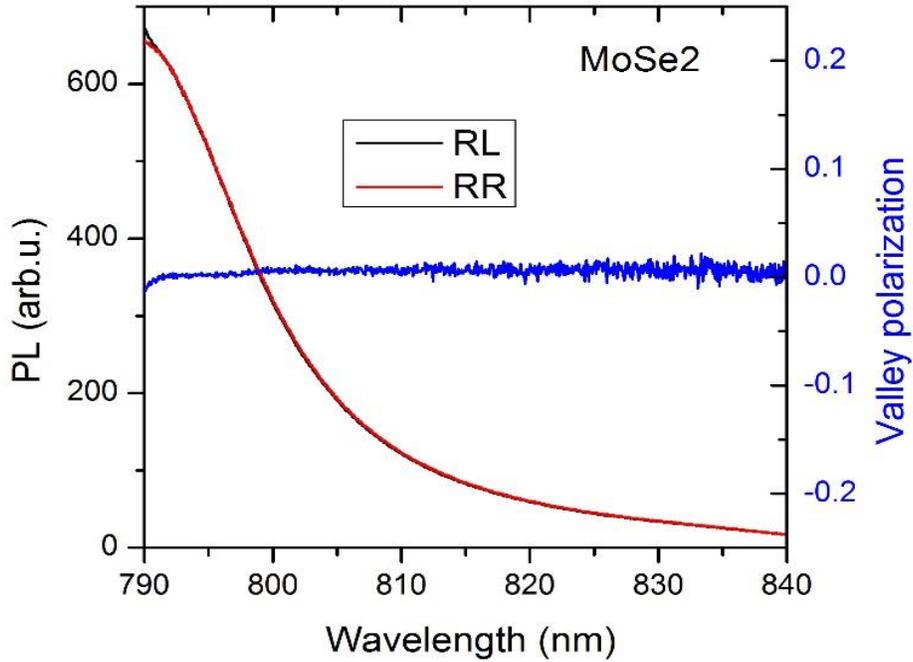

**Figure S3. Circular polarized PL of exfoliated monolayer MoSe$_2$ and the calculated degree of valley polarization. The measurements are carried out with a continuum laser at 780 nm. RL: The excitation is induced by a right circularly polarized laser and detection is selected to the left circularly polarized PL (black line), RR: the excitation is induced by a right circularly polarized laser and the right circularly polarized PL is detected (red line). Blue line is the degree of valley polarization calculated according to: P = (RR-RL) / (RR+RL).**

## 4) Estimation of the exciton dephasing time

At room temperature, the linewidth of the exciton given by the PL spectra is around 40 meV. This finite width has its origin in the contribution of both the inhomogeneity and homogeneity broadening. For the inhomogeneity part, we note that at very low temperature (~10 K), the width of the measured PL is around 10 meV,[5] which we take as an approximation of the inhomogeneity width. The homogeneity broadening line width $\hbar\Delta v$ at room temperature is estimated to be around 38.73 meV by taking the deconvolution, which gives an exciton de-phasing time $\tau \approx (\pi\Delta v)^{-1} = 34$ fs due to thermo effect induced exciton-phonon and exciton-exciton collision.